\documentclass[aps,prl,a4paper,twocolumn,superscriptaddress,citeautoscript]{revtex4-1}
\usepackage{color}
\usepackage[english]{babel}
\usepackage[T1]{fontenc}
\usepackage{amsmath}
\usepackage{hyperref}
\usepackage{units}
\usepackage[charter]{mathdesign}
\usepackage[pdftex]{graphicx}
\usepackage{color}
\begin{document}
\title{Impact of antiferromagnetism on the optical properties of rare earth nickelates}
\author{J.~Ruppen}
\affiliation{Department of Quantum Matter Physics, University of Geneva, 24 
Quai Ernest-Ansermet, 1211 Geneva 4, Switzerland}
\author{J.~Teyssier}
\affiliation{Department of Quantum Matter Physics, University of Geneva, 24 
Quai Ernest-Ansermet, 1211 Geneva 4, Switzerland}
\author{I.~Ardizzone}
\affiliation{Department of Quantum Matter Physics, University of Geneva, 24 
Quai Ernest-Ansermet, 1211 Geneva 4, Switzerland}
\author{O.~E.~Peil}
\affiliation{Department of Quantum Matter Physics, University of Geneva, 24 
Quai Ernest-Ansermet, 1211 Geneva 4, Switzerland}
\affiliation{Coll\`ege de France, 11 place Marcelin Berthelot, 75005 Paris, 
France}
\author{S.~Catalano}
\affiliation{Department of Quantum Matter Physics, University of Geneva, 24 
Quai Ernest-Ansermet, 1211 Geneva 4, Switzerland}
\author{M.~Gibert}
\affiliation{Department of Quantum Matter Physics, University of Geneva, 24 
Quai Ernest-Ansermet, 1211 Geneva 4, Switzerland}
\author{J.-M.~Triscone} 
\affiliation{Department of Quantum Matter Physics, University of Geneva, 24 
Quai Ernest-Ansermet, 1211 Geneva 4, Switzerland}
\author{A.~Georges}
\affiliation{Department of Quantum Matter Physics, University of Geneva, 24 
Quai Ernest-Ansermet, 1211 Geneva 4, Switzerland}
\affiliation{Coll\`ege de France, 11 place Marcelin Berthelot, 75005 Paris, France}
\affiliation{Centre de Physique Th\'eorique, \'Ecole Polytechnique, CNRS, 91128 
Palaiseau Cedex, France}
\author{D.~van der Marel}\email{dirk.vandermarel@unige.ch}
\affiliation{Department of Quantum Matter Physics, University of Geneva, 24 
Quai Ernest-Ansermet, 1211 Geneva 4, Switzerland}
\date{\today}

\date{\today}
\begin{abstract}
We study the temperature dependence of the optical conductivity of rare earth nickelate films of varying composition and strain close to the antiferromagnetic ordering temperature,  $T_N$. Two prominent peaks at 0.6 and 1.3 eV that are characteristic of the insulating phase, display a {small but} significant increase in intensity when the material passes from para- to antiferromagnetic. This observation indicates the presence of a positive feedback between antiferromagnetic {(AF) and bond disproportionation (BD)} order. By analyzing the temperature dependence near $T_N$, and using a Landau-type free energy expression for {BD and AF} order, we infer that {BD} order is a necessary condition for the AF phase to appear, and that the antiferromagnetism contributes to stabilization of the {bond disproportionation}. This model also explains why hysteresis is particularly strong when the transition into the insulating state occurs simultaneously with antiferromagnetic order.

\end{abstract}
\maketitle
\begin{figure}[t!!]
\begin{center}
\includegraphics[width=1.0\columnwidth]{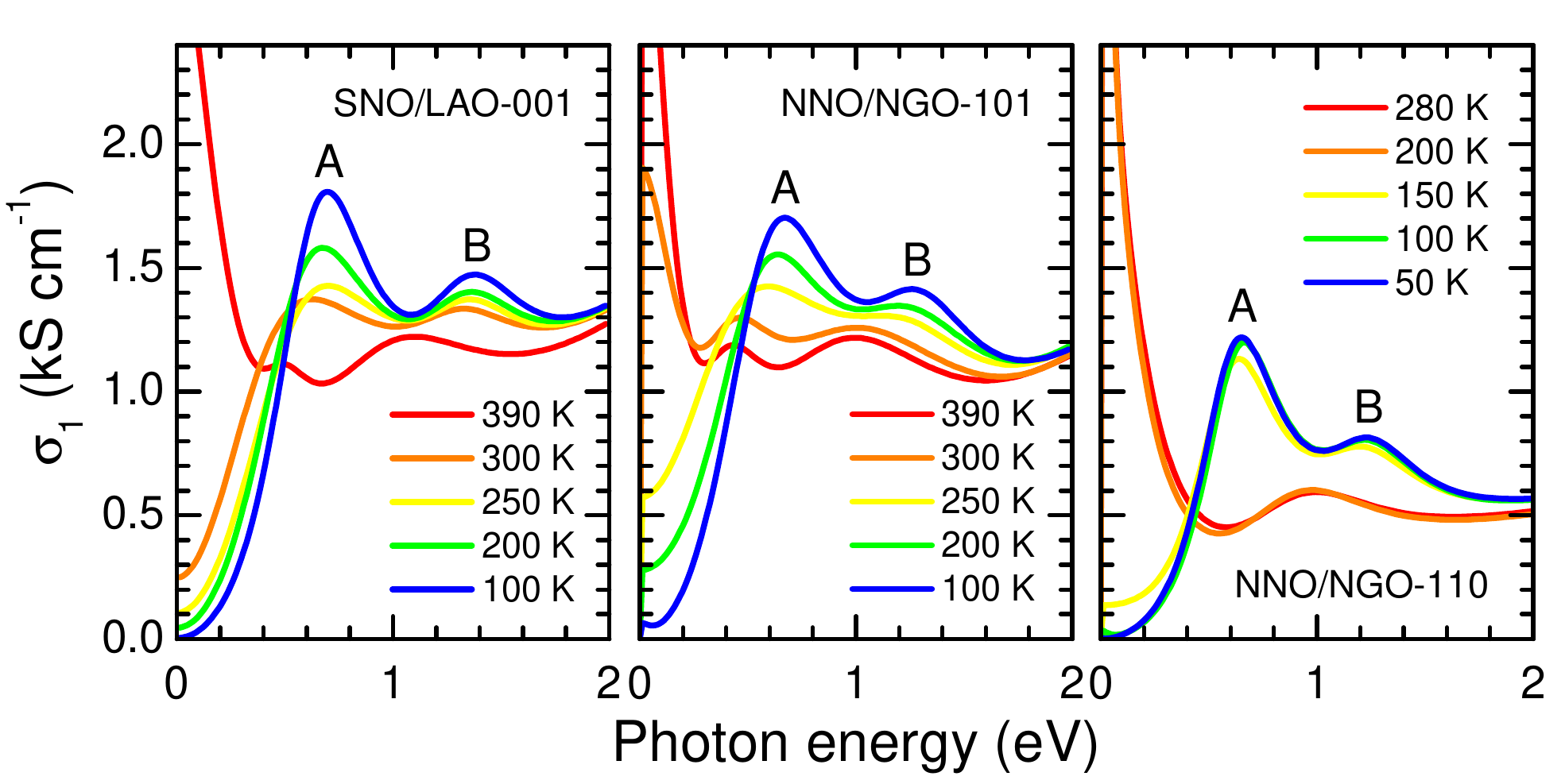}
\caption{\label{fig:conductivity}
Optical conductivity spectra for selected temperatures of SmNiO$_3$ on a LaAlO$_3$ $(001)$ substrate, NdNiO$_3$ on NdGaO$_3$(101), and NdNiO$_3$ on NdGaO$_3$(110)~\cite{ruppen2015}. }
\end{center}
\end{figure}

%
%
Rare earth nickelates form a class of transition metal oxides that undergo a metal-insulator transition as a function of the temperature $T$, and the so-called tolerance factor, $t$, which describes the distortion of the crystal structure associated to tilting and rotations of the oxygen octahedra surrounding the Ni-atoms.
Depending on $t$ that can be tuned by rare earth radius or strain, the material (i) remains metallic for all temperatures, (ii) switches in a first order phase transition to an antiferromagnetic insulator at $T_{MI}$, or (iii) traverses two phase transitions, the highest one at $T_{MI}$ being from metal to paramagnetic (PM) insulator, and the lowest one being the N\'eel temperature $T_N$ where the material becomes antiferromagnetic (AF). 
The insulating phase of rare-earth nickelates is understood in terms of inequivalent nickel sites. In an extreme picture, every second nickel site is in a $d^8$ configuration and carries a magnetic moment while the other ones are in a non-magnetic $d^8\underline{L}^2$ configuration~\cite{alonso1999,mizokawa2000,park2012,lau2013,johnston2014,subedi2015}.
Due to electron-lattice coupling the long-range charge order is accompanied by a breathing lattice distortion~\cite{mazin2007,chaloupka2008} opening a Peierls gap in the energy range 0.5-0.7 eV above the Fermi energy~\cite{subedi2015,ruppen2015}. 
{We refer to this modulation as bond disproportionation (BD) order}.
The magnetically ordered phase is characterized by a wave-vector \textbf{k} = (1/4,1/4,1/4)~\cite{rodriguez1998} in  pseudocubic notation. Two magnetic structures were proposed to explain the magnetic origin of this diffracted intensity: up-up-down-down~\cite{rodriguez1998} and non collinear ordering~\cite{scagnoli2006}. 
More recent measurements confirm the non-collinear structure~\cite{scagnoli2008}. 
The relationship between the {BD} and the AF order is still under debate.
The optical conductivity in the insulating phase of RNiO$_3$ is characterized by two strong peaks at 0.6 eV and 1.3 eV (peaks A and B respectively)~\cite{stewart2012,ruppen2015,toriss2017} (Fig.~\ref{fig:conductivity}). In a recent paper we have reported the changes of the optical conductivity spectrum of SmNiO$_3$ and NdNiO$_3$ films~\cite{ruppen2015}, and compared the spectra in the metallic and insulating states. 
The aforementioned features of the optical spectra were found to be well reproduced by Dynamical Mean Field Theory (DMFT) calculations, allowing the identification of two peaks at 0.6 and 1.3 eV as the transitions across the Mott-insulating gap and the Peierls pseudogap respectively. The optical conductivity in the metallic phase is characterized by  a zero-energy mode and a peak at 1 eV. 

Here we report on the impact of AF order on the optical conductivity spectrum. We observe that antiferromagnetism in SmNiO$_3$ and NdNiO$_3$ thin films strongly influences peaks A and B, and causes an enhancement of the oscillator strength of these two peaks. The observed temperature dependence corresponds to a soft onset at the N\'eel temperature, signaling a positive feedback between AF and {BD} order. Hysteresis of the optical spectra is strong when AF and {BD} order occurs simultaneously, and negligible when these occur at separate temperatures, consistent with Refs.~\onlinecite{granados1992,medarde1997,vobornik1999,catalan2008,kumar2009,caviglia2012,scherwitzl2010}. 
%
%
%
\begin{table}[]
\begin{ruledtabular}
\begin{tabular}{lclclclclclclcl}
Film/Substrate&ps-c& Thickness & Strain & $T_{MI}$&$T_{N}$ \\
\hline
SNO/LAO-001&001& 10 nm & -0.1\% & 380 K& 200 K \\
NNO/NGO-101&111& 17 nm & +1.5\% & 335 K& 240  K \\
NNO/NGO-110&001& 30 nm & +1.5\% & 160 K& -
\end{tabular}
\end{ruledtabular}
\caption{\label{table:1} Film/substrate properties. First and second column give standard and pseudo-cubic substrate orientation respectively. $T_{MI}$ of NNO/NGO-101 is much higher than of NNO/NGO-110 due to the specific tilts of the oxygen octahedra of the former~\cite{catalano2015}.}
\end{table}
We analyze thin films of SmNiO$_3$ on LaAlO$_3$(001)~\cite{catalano2014} and NdGaO$_3$ substrates~\cite{catalano2015} labeled SNO/LAO-001, NNO/NGO-101 and NNO/NGO-110, the properties of which are summarized in table~\ref{table:1}. 
The optical conductivity (Fig.~\ref{fig:conductivity}) was measured as described in Ref.~\onlinecite{ruppen2015}. 
The metal insulator transition is revealed in the optical conductivity as the loss of the zero-frequency mode and the appearance of the peaks at 0.6 eV (peak A) and 1.3 eV (peak B). 
\begin{figure}[ht!]
\begin{center}
\includegraphics[width=\columnwidth]{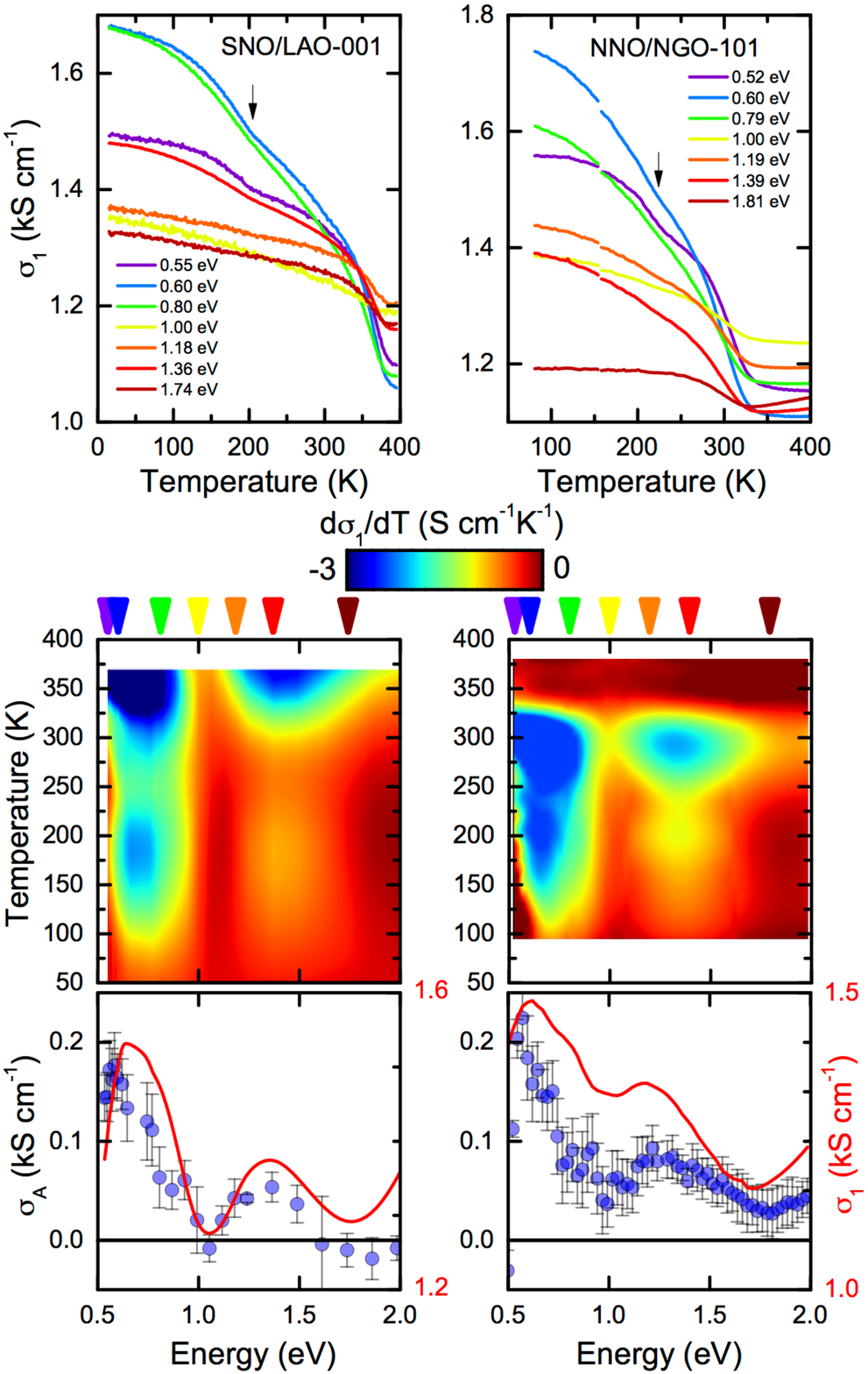}
\caption{\label{fig:kink}
Temperature dependence of the optical conductivity, $\sigma_1(\omega,T)$, at selected photon energies for (top left) SNO/LAO-001 and (top right) NNO/NGO-101. The small jump at 150 K in the NNO/NGO-101 data is an experimental artifact and is ignored in the discussion. Middle panels: color plots of $d\sigma_1(\omega,T)/dT$. Bottom panels: The optical conductivity spectra at the N\'eel temperature, $\sigma_1(\omega,T_N)$ (red curves), and the antiferromagnetism induced contribution to the optical conductivity, $\sigma_{A}(\omega)$ (blue circles).   }
\end{center}
\end{figure}

To highlight the temperature dependence through $T_{MI}$ and $T_{N}$ we show in Fig.~\ref{fig:kink} the temperature dependence of the optical conductivities of  SNO/LAO-001 and NNO/NGO-101 for selected photon energies. 
Also shown are color plots of $d\sigma_1(\omega,T)/dT$ in the frequency-temperature plane. Most clearly visible in these data is the metal-insulator transition at $T_{MI}$. However, for some of the photon energies there is a soft step at temperatures $T_i$ close to the N\'eel temperature (highlighted by arrows in Fig.~\ref{fig:kink}). The softness of these steps is an important feature helping us to understand better the nature of the coupling between {BD} and AF order, to which we will return below. For now we notice that the strength of the steps of $\sigma_1(\omega,T)$ around $T_i$ seems to correlate with peaks A and B. In  $d\sigma_1(\omega,T)/dT$ this shows up as the maxima at $T_i$ for peak A and peak B.

We are interested in the additional conductivity spectrum, $\sigma_{A}(\omega)$, arising from the AF order. Leaving the softness of the step at $T_i$ for discussion later in this article, we fitted for each photon energy a polynomial of the form ${s_0}^{\pm}+{s_1}^{\pm} (T-T_i) +{s_2}^{\pm} (T-T_i)^2$ to  $\sigma_1(\omega,T)$ in a broad temperature range (about 100 K)  above ($+$) and below ($-$) the inflection point $T_i$. The quantity $\sigma_{A}(\omega)=[{s}_{1}^{+}(\omega)-{s}_{1}^{-}(\omega)]T_i/2$ then represents, apart from a factor of order one, the additional conductivity  spectrum extrapolated to zero temperature. 
The results of this analysis are shown in the bottom panel of Fig.~\ref{fig:kink}. We see, that this corresponds to a reinforcement of the double peak structure already present in the PM insulating phase. 

\begin{figure}[t!!!]
\begin{center}
\includegraphics[width=\columnwidth]{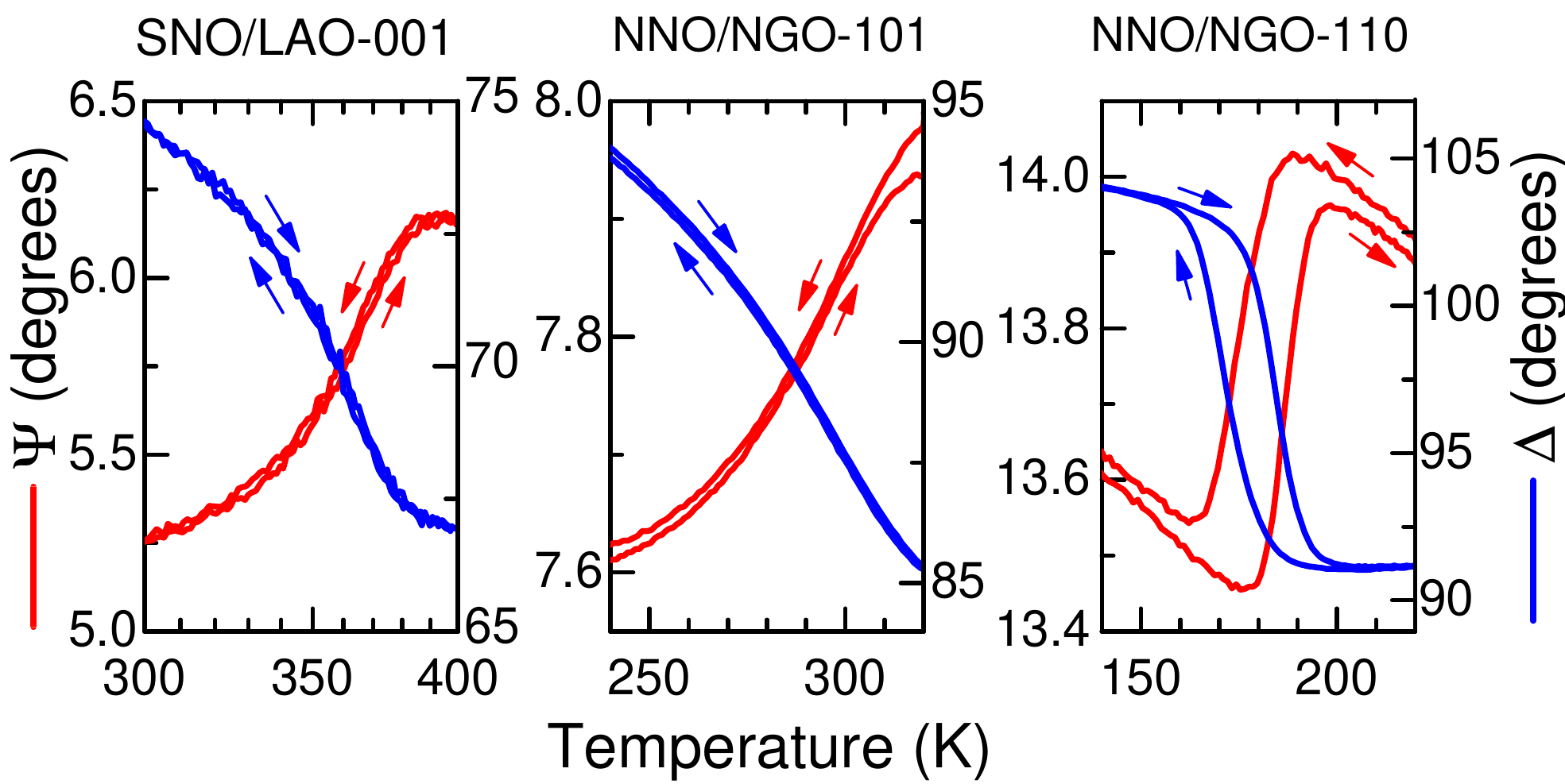}
\caption{\label{fig:hysteresis}
Temperature dependence of the ellipsometric parameters of the three samples, showing strong hysteresis for  NNO/NGO-110 (right panel), and absence of hysteresis for NNO/NGO-101 (central panel) and SNO/LAO-001 (left panel). Red curves are during warm-up, blue curves during cool-down. }
\end{center}
\end{figure}

We now turn to an experimental observation that is of crucial importance for the subsequent discussion: As pointed out in the introduction, we can directly associate peaks A and B with spectral features reflecting the {bond disproportionation}, which are absent in the metal phase. The extra spectral weight in the two peaks as the temperature is lowered below $T_N$ indicates that the {BD} order is enhanced in the AF state. We moreover see, that the effect of AF order is by and large limited to an increase of the intensity of peaks A and B. The intensity at or near the peak position, is therefore a measure of the order parameter $\Phi$ characterizing the {BD} order. More precisely, since the optical spectra and free energy are insensitive to the sign of $\Phi$, we associate this intensity, apart from a temperature independent background contribution, to $\Phi^2$. 
To analyze what happens at $T_N$ we follow a phenomenological approach employing the Landau theory of phase transitions, where
the antiferromagnetism is characterized by the order parameter $m$. The free energy is an even function of both $m$ and $\Phi$, and can be expanded as follows~\cite{imry1975,footnote,lee2011}
\begin{equation}
    f = {a}\Phi^2 + \frac{b}{2 !}\Phi^4 + \frac{c}{3 !}\Phi^6+ {\lambda} \Phi^2 m^2 + \alpha m^2+\frac{\beta}{2 !} m^4 
\label{eq:f}    
\end{equation} 
{The Landau theory is a fairly general framework for analyzing the coupled effects of {BD} and AF ordering, applicable to different situations not explicitly considered in our paper ({\em e.g.} heterostructures).  However, the detailed dependence of the coefficients of the Landau functional on the control parameters at hand  will have to be considered in a case by case manner. Here we focus on the pristine RNiO$_3$ compounds, and we will assume a  specific dependence of these coefficients on a single parameter characterizing the distortion (due to strain, rare-earth substitution, etc). This is  mostly for illustrative purposes, and should not be taken as an indication that all situations of interest can be analyzed using this single  parameter. However, the form of the Landau functional should have a larger degree of validity. 

The generic behavior of the nickelates is, that no antiferromagnetism occurs in the metal phase~\cite{superlattices,frano2013}. This indicates that  {BD} is necessary for AF to occur, implying that $\alpha=0$. From here on we will use this condition on $\alpha$ as a hypothesis, which will be justified {\em a posteriori} by the temperature dependence of our optical data in the AF phase.}
We will assume that the main temperature dependence close to the magnetic and metal-insulator transitions enters through ${a}$ and ${\lambda}$ 
\begin{eqnarray}
  {a}(T) &=& {a}_0 \left( [T/T_{MI}]^{\eta} -1   \right) \nonumber \\
  {\lambda}(T) &=& {\lambda}_0 \left( [T/T_{N}]^{\eta}  -1 \right)
\label{eq:ad}   
\end{eqnarray}
The AF and {BD} order parameters $m(T)$ and $\Phi(T)$ saturate at low temperatures, which we qualitatively describe by $\eta=4$ (the precise value of $\eta$ is not essential to our arguments). Below the temperatures $T_{MI}$ ($T_{N}$) the coefficient ${a}$ (${\lambda}$) becomes negative.  $T_{MI}$ and $T_{N}$ are material parameters 
{which for simplicity will be assumed to depend on a single parameter, namely}
the tolerance factor $t$, for which we will use the phenomenological parameterization
%
\begin{eqnarray}
  T_{N}(t) &=&  {\theta}_{N}(t-t_{N})  \nonumber \\
  T_{MI}(t) &=&  {\theta}_{MI}(t_{MI}^+-t)(t-t_{MI}^-)
\label{eq:TCOTN}   
\end{eqnarray}
with the parameters given in Table \ref{table:2}. Since we are not interested in the absolute values of $m$, $\Phi$ and $f$, we choose the scale of these quantities such as to provide $c=\beta={a}_0=1$. This leaves ${b}$ and ${\lambda}_0$ as the only adjustable coefficients of our model. Finally we note, that several experiments indicate that for large values of the tolerance factor there is a single first order transition, whereas at least in part of the phase diagram where there are two phase transitions, the AF transition is second order and the metal-insulator transition weakly first order. The latter could indicate that the metal-insulator transition is {\em a priori} second order, while driven first order by coupling to the lattice.  The splitting of a single first order transition into a set of second order transitions requires that ${b}=0$, a fine tuning that is unlikely to occur by chance. Since several experiments have indicated first order behaviour at $T_{MI}$ for most, if not all, values of the tolerance factor, we will choose ${b}=-0.25$ for the coefficient associated to {bond disproportionation}, and  ${\lambda}_0=0.8$ for the $\Phi^2 m^2$ coupling. The negative value of the parameter ${b}$ (implying a first order transition at $T_{MI}$ everywhere in the phase diagram) is to be understood as a consequence of the positive feedback of the electron-lattice coupling on the {bond disproportionation}. As we will see, this choice of parameters illustrates qualitatively several thermodynamic aspects of these materials, including the temperature dependence of the optical experiments reported in the present manuscript.
\begin{table}[ht]
\begin{tabular}{|c |c |c |c | c| c|  }
\hline
${a}_0$&  ${b}$ & ${c}$& ${\lambda}_0$ & ${\alpha}$&$ {\beta}$     \\
1&  -0.25 & 1 & 0.8 & 0& 1    \\
\hline
${\theta}_{N}$ (K) & ${t}_{N}$ &${\theta}_{MI}$ (K)&  ${t}_{MI}^+$& ${t}_{MI}^-$ &  $\eta$ \\
2597 & 0.804&101198&  0.927 & 0.773&  4 \\
\hline
\end{tabular}
\caption{\label{table:2} Parameters used for the calculations in Figs. \ref{fig:model} and \ref{fig:model_kink}. {The parameters on the second lign were chosen such as to mimic in a coarse graining manner the evolution of $T_{MI}$ and $T_N$ as a function of tolerance factor reported in Ref.~\cite{catalan2008}}.}
\end{table}
\begin{figure}[ht!]
\begin{center}
\includegraphics[width=\columnwidth]{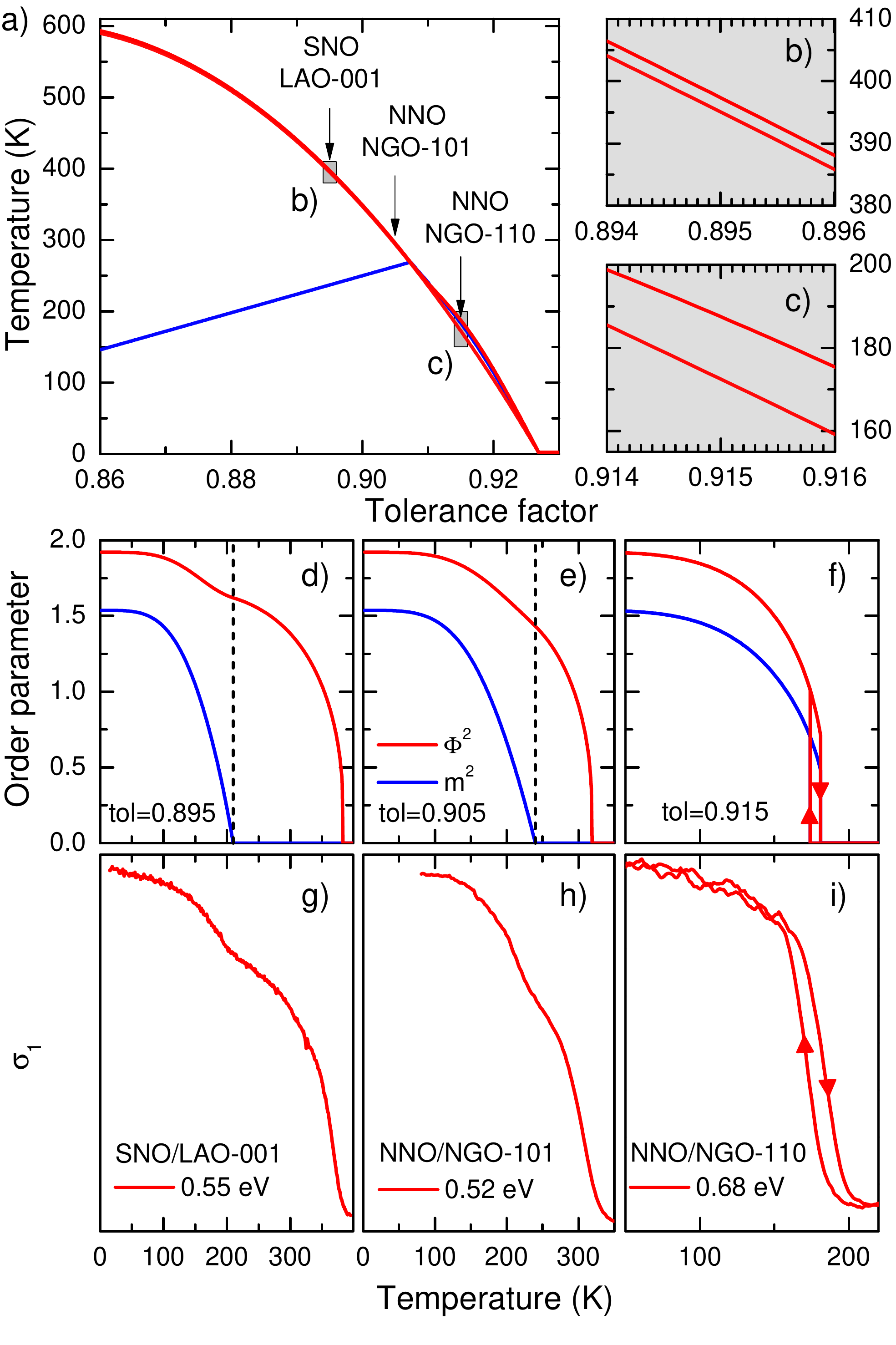}
\caption{\label{fig:model}
a) Calculated phase diagram in the temperature / tolerance factor plane of RNiO$_3$. Hysteresis limit curves enlargement centered on SNO/LAO-001 and on NNO/NGO-110 are shown in b) and c) respectively. d-f) Temperature dependence of the AF ($m^2$) and metal insulator ($\Phi^2$) order parameters for three selected values of the tolerance factor, such as to match the experimental metal-insulator transition temperatures of g) SNO/LAO-001, h) NNO/NGO-101 and i) NNO/NGO-110.}
\end{center}
\end{figure}
The phase diagram in the tolerance factor / temperature plane can now be reproduced with the help of the parameters described above and with Eq. \ref{eq:f}. The result is displayed in Fig.~\ref{fig:model}a). In the region of hysteresis the free energy has a metastable minimum coexisting with the stable one. Interestingly the phase diagram shows a widening of this region around $t=0.915$ for the following reason: The free energy (Eq. \ref{eq:f}) at its minimum with respect to $m$ equals
\begin{eqnarray}
    f &=& {a}\Phi^2 + \frac{\tilde{b}}{2 !}\Phi^4 + \frac{c}{3 !}\Phi^6\\
    \tilde{b}&=&b-\lambda^2 \hspace{2mm}(T<T_N)\hspace{5mm};\hspace{5mm} \tilde{b}=b\hspace{2mm}(T>T_N) \nonumber
    \label{eq:f2}        
\end{eqnarray} 
Since a first order transition requires a negative value of the coefficient $\tilde{b}$, the $-\lambda^2$ contribution to $\tilde{b}$ enhances the first order character and the size of the hysteresis loop around $T_{MI}${.}
The hysteresis of  about 15 K is close to the behavior observed in the bulk compound NdNiO$_3$ (and sample NNO/NGO-110). Such hysteresis is also present in the optical spectra, shown in Fig.~\ref{fig:hysteresis} for the original ellipsometric parameters $\Psi$ and $\Delta$ from which the optical conductivity was obtained using the method described in Ref.~\onlinecite{ruppen2015}. Since these thermal cycles take several hours, the small differences observed in SNO/LAO-001 and NNO/NGO-101 during heating and cooling may be partly or entirely caused by instrument drift or absorption and desorption of a small quantity of gas molecules at the sample surface. 

In the middle panels of Fig.  \ref{fig:model} the temperature dependence of $\Phi^2$ and $m^2$ calculated for $t=0.895$ (SNO/LAO-001, Fig.~\ref{fig:model}b), $t=0.905$ (NNO/NGO-101, Fig.~\ref{fig:model}c), and $t=0.915$  (NNO/NGO-110, Fig.~\ref{fig:model}d) is shown. While for all cases the calculation shows that the transition at $T_{MI}$ is first order, the hysteresis of the two formers are too small to display on this scale whereas, for the latter, the hysteresis is about 10 K, and shows clearly in the temperature dependence. The experimental data of the optical conductivity at an energy close to peak A, shown in Fig.~\ref{fig:model}e-g, closely follow these trends.
\begin{figure}[t!!!]
\begin{center}
\includegraphics[width=\columnwidth]{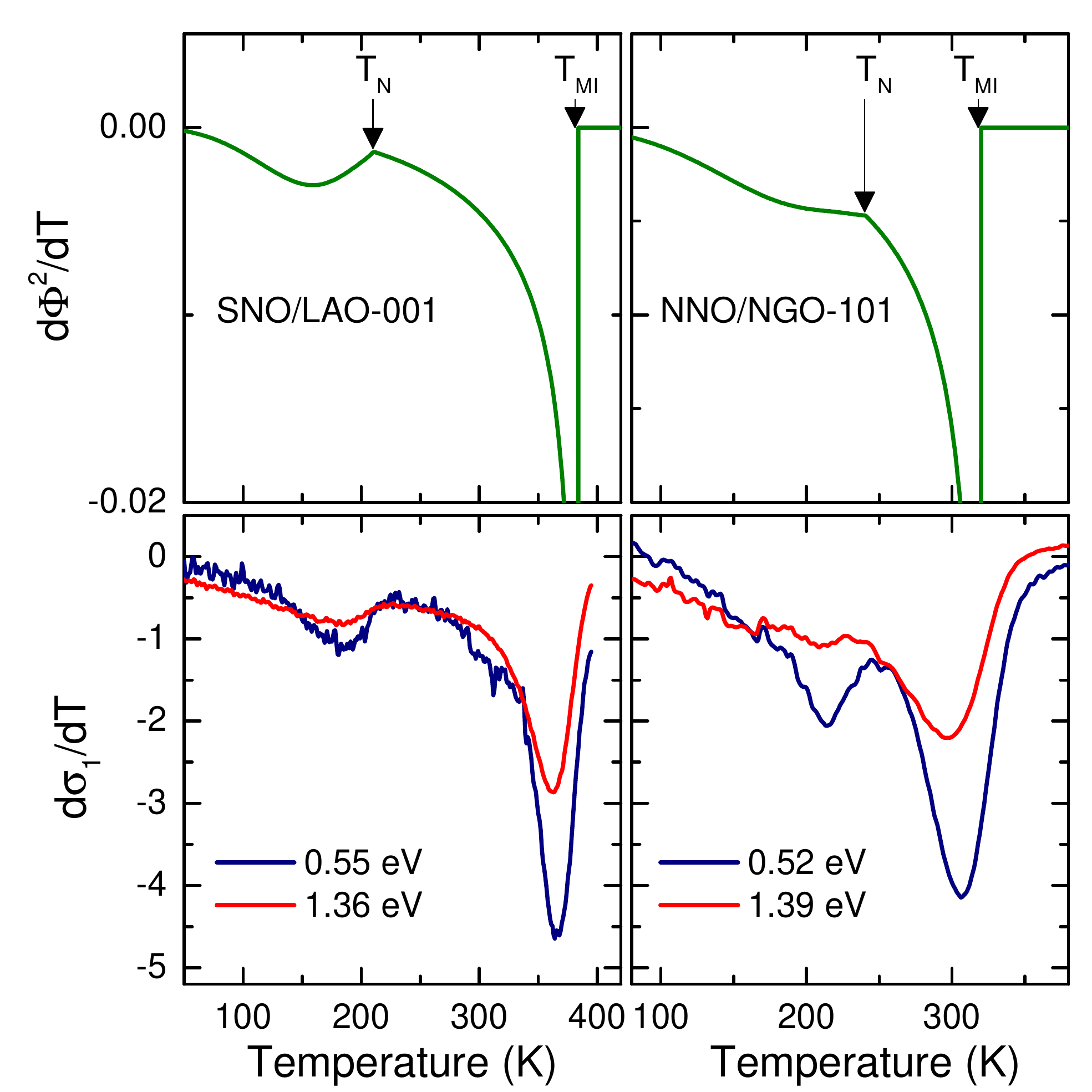}
\caption{\label{fig:model_kink}Temperature derivative of the {BD} parameter $\Phi^2$ (top panels) and the temperature derivative of the optical conductivities of SNO/LAO-001 and NNO/NGO-101 (bottom panels) measured at photon energies corresponding to the maximum of peaks A (navy) and B (red). }
\end{center}
\end{figure}
We now turn our attention to the details of the temperature dependence close to $T_N$.  If the intensity of peaks A and B were to track the antiferromagnetic order, a kink would be expected at $T_N$. The experimental data in the top panel of Fig.~\ref{fig:kink} and bottom panel of Fig.~\ref{fig:model}  show only an inflexion point, not a kink. In principle it is not unusual that some broadening occurs, for example if $T_N$ would not be completely homogeneous across the area of the sample. The kink in $\sigma_1(\omega,T)$ would then be replaced by an inflection point, and one would be tempted to associate the average N\'eel temperature with this inflection point. However, the phenomenological Landau modeling tells a different story: Minimizing $f(\Phi,m)$ (Eq. \ref{eq:f}, $c=\beta=1$) with respect to $\Phi$ and $m$ and eliminating ${\lambda}$, leads to the following relation between the equilibrium values of $\Phi^2$ and $m^2$ 
\begin{equation}
\Phi^2=-{b}\pm\sqrt{{b}^2-2{a}+2{m^4}/\Phi^{2}}
     \label{eq:Phi2}    
\end{equation} 
Deep inside the insulating state and close to $T_N$ we can expand the right hand side in $m^4/\Phi^{2}$. The leading order of this expansion is proportional to $m^4$, which close to the transition is proportional to $(T_N-T)^2$. This soft onset of the AF-induced contribution to $\Phi^2$ is a direct consequence of {our hypothesis that $\alpha=0$ as supported by our data of $d\sigma_1/dT$ exhibiting a kink and not a jump at $T_N$ as shown in the bottom panel of Fig.~\ref{fig:model_kink} for samples  SNO/LAO-001 and NNO/NGO-101. Comparing this to the theoretical $d\Phi^2/dT$ (top panel of the same figure) we conclude that, with the parameters ${\lambda}_0=0.8$, ${b}=-0.25$, and $\alpha=0$ the temperature trend of the spectral weight and its behaviour at the phase transitions, is well described by the Landau-theory. 
Note also that, as a result of the soft onset, the inflection point in the $\Phi^2(T)$ curve occurs well below the actual $T_N$.  The requirement that $\alpha=0$  corroborates another experimental observation, namely that these compounds are paramagnetic in the metal phase. }

We observed a {small but} significant impact of antiferromagnetic (AF) order on the optical conductivity spectrum of RNiO$_3$. The intensity of two prominent conductivity peaks was previously demonstrated to track the charge order accompanied by bond disproportionation (BD) in these compounds~\cite{ruppen2015}. We now observe that in the antiferromagnetic state an additional spectral weight is added, proportional to $m^4$ where $m$ is the antiferromagnetic order parameter. This soft onset of the AF-related spectral weight proves that the BD is a {\em conditio sine qua non} for the AF order, and is excellently described by a Landau model for the free energy with two coupled (BD and AF) order parameters. The temperature dependence upon thermal cycling indicates that the transition into a simultaneously {BD} and AF ordered phase, has much stronger hysteresis than the transition into a BD phase without AF order. This aspect is also well described by the aforementioned Landau model. These observations and conclusions permit to describe a wealth of transport and spectroscopic data in a unified thermodynamic framework, using a small set of Landau parameters that may serve as a basis of future microscopic models.
\begin{acknowledgments}
We gratefully acknowledge discussions with G.~A.~Sawatzky, G.~Khaliullin, D.~I.~Khomskii, M.~Medarde and I.~I.~Mazin. This project was supported by the Swiss National Science Foundation (projects 200020-165716, 200021-146586 and NCCR MARVEL) and by the European Research Council (ERC 319286-QMAC).

J.R. and J.T. contributed equally to this work.
\end{acknowledgments}

\end{document}